# Some Remarks on Nonlinear Properties of Pumping System of an Optical-Wavelength Acoustic Laser (Phaser)

## D. N. Makovetskii

Institute of Radio-Physics and Electronics of National Academy of Sciences,
12, Academician Proskura Street, Kharkov, Ukraine, 61085
e-mail: dmakov@rambler.ru

## Abstract

New critical surfaces have been analytically found for a 3D vectorial model of bistable pumping system of an optical-wavelength acoustic laser (phaser). The Lyapunov instability is possible for this pumping system at the upper branch for a high quality factor $Q_c$ of a pump resonator: $Q_c >> Q_m$, where $Q_m$ is the magnetic quality factor.

## 1. Introduction

The problem of realization of stable stationary inversion states in optical-wavelength acoustic lasers (phasers) [19–22] is very complicated for the millimeter-range pumping frequencies. The main difficulties are caused by the pronounced shortening of longitudinal spin relaxation time $\tau_{1p}$ at pump frequency. The shorter is $\tau_{1p}$, the higher must be amplitude $H_{1p}$ of the input microwave pump field. On the other hand, the typical cryogenic phaser systems can not rich pump power of order 1 W without active crystal overheating. The only way to excite high-amplitude field in active crystal by power-limited pump is the using of good quality $Q_c >> 1$ resonance structures, e.g. microwave Fabry-Perot resonators (MFPR) or microwave whispering-gallery resonators (MWGR).

Because of paramagnetic medium nonlinearity the microwave re-emission of pump field by active centers in resonance structures leads to forming of multiattractor states of dissipative system under consideration [1]. Firstly such the phenomenon for active microwave system of paramagnetic quantum amplifiers was predicted theoretically in [2, 3] and observed experimentally [4, 8] andalusite quantim amplifier with $Fe^{3+}$ active centers. There is now theoretical and experimental evidence of bistability in medium-quality dissipative pump systems. But in high-quality MFPR, namely at $Q_c >> Q_m$ ($Q_m$ is the magnetic quality factor [5]) the stability properties of dissipative pump system are not investigated. On the other hand, quality factors of pump resonators in optical-wavelength phasers are high [19–22]. So the problem of stability of resonator pumping in phasers has an undoubted interest.

## 2. Mechanisms of nonlinearity in phaser pump systems

The possibility for observing the nonlinear processes in solid-state quantum amplifiers has been originally discussed as long ago as in 1960s (without any attempts to study stability of inversion states) [5–7]. Yet, the nonlinearity parameter $\beta_N$ (which had been found in these works) turned out to be equal to the product of the pumped spin transition saturation factor $Z_p$ into the relation parameter $r_{21} \equiv \tau_{2p}/\tau_{1p}$, where $\tau_{2p}$ is the transverse spin relaxation time. Unlike the gaseous masers and optical lasers active media where the relation parameter $r_{21}$ can be of the order of 1, the values $r_{21} = 10^{-4} - 10^{-8}$ are typical for the dilute paramagnets being used in the phaser amplifiers. Such circumstance makes it difficult to bring out the phenomena predicted in [5–7] since it is hard to achieve the saturation both with great $Z_p$ and without paramagnetic crystal overheating, — however it is necessary to obtain the values $\beta_N$ on the order of 1. On the other hand, the experiments on the real microwave phaser amplifiers showed that the appreciable deviations from linearity was actually observed even with $\beta_N \ll 1$, which suggests that a qualitatively different mechanism of nonlinearity is predominant here. This mechanism will not be associated with the need for so deep spin-system saturation ($Z_p \approx 10^4 - 10^8$), as it is required by the model proposed in [5–7].

A model of resonant paramagnetic dissipative system saturation, involving the self-action of the saturating field through a spin-system in a electromagnetic MFPR was developed in [2, 3, 8] for the cases of $q_1 \equiv 2\tau_c/\tau_{1p} \ll 1$ and $q_1 \approx 1$, where $\tau_c$ is the photon lifetime in an empty MFRP. In contrast to the model reported in [5-7] the nonlinearity parameter in [2, 3, 8] is derived as $\xi \equiv q_2 \xi_0 = Q_c^{(0)}/Q_m$, where $q_2 \equiv 2\tau_c/\tau_{2p}$; $\xi_0$ is independent on $q_2$ value; $Q_c^{(0)}$ is loaded quality factor of MFPR outside the magnetic resonance domains. As it was found in [2, 3, 8] with $\xi > 1$ the particularly pronounced nonlinear phenomena such as stationary states branching, instabilities, automodulation etc. may arise in dissipative paramagnetic systems even at $Z_p \approx 10 - 10^2$ as resulting due to strong internal spin-photon feedback in the MFPR. Thus, the applicability conditions of the nonlinear model [2, 3, 8] are in good consistency with the real values of control parameters in the microwave quantum amplifiers.

## 3. Equations of motion

The shortening of $\tau_{1p}$ in millimeter range do not change the inequality $\tau_{1p} \gg \tau_{2p}$ at liquid helium temperatures, but quality factor of MFPR needed for effective phaser pumping will raise in mentioned range to values, for which resonant linewidth of MFPR $\Delta_c \propto 1/\tau_c$ may be of the same order as linewidth of electron paramagnetic resonance (EPR) $\Delta_m \propto 1/\tau_{2p}$ for spin pump transition of active medium or even much smaller than the EPR linewidth. The optimal relationship between $\tau_c$ and $\tau_{2p}$ with which the maximum of inversion ratio $K$ is realized appears to be highly essential. It is necessary at this point to clarify the role of $q_2$ as the *control parameter* of our system. From this point of view the flow-type [9, 10] equations of motion must have the form:

$$\frac{dX_i}{dt} = \Psi_i^{(L)}(X_j; q_2, P, \ldots) + \Psi_i^{(NL)}(X_j; q_2, P, \ldots), \qquad (1)$$

where $X_i, X_j$ are the components of vectorial order parameter $\mathbf{X}$; $\Psi_i^{(L)}$ and $\Psi_i^{(NL)}$ represent linear and nonlinear parts of flow in phase space; $P$ is another control parameter (independent on $q_2$), describing intensity of pumping. The full set of control parameters includes not only $q_2$ and $P$, but has dimension of 5 – 7 for real phaser system. All the control parameters denoted in (1) as triple period will be defined later.

Let us reduce the general Maxwell-Bloch equations of motion (see, e.g book [1]) for nonlinear electromagnetic MFPR with paramagnetic centers to the flow equations (1). For Fe-group paramagnetic centers Maxwell-Bloch equations set reads:

$$\left. \begin{array}{l} \dfrac{\partial \rho_{\mu\mu}}{\partial t} = -\dfrac{i}{\hbar}\left[\hat{H};\hat{\rho}\right]_{\mu\mu} - \rho_{\mu\mu}\sum_{\nu}W_{\mu\nu} + \sum_{\nu}\rho_{\nu\nu}W_{\nu\mu} \\ \dfrac{\partial \rho_{\mu\nu}}{\partial t} = -\dfrac{i}{\hbar}\left[\hat{H};\hat{\rho}\right]_{\mu\nu} - \rho_{\mu\nu}/\tau_{2,\mu\nu}; \quad \mu \neq \nu \end{array} \right\}; \qquad (2a)$$

$$\dfrac{\partial^2 \tilde{H}_1}{\partial t^2} - V_e^2 \dfrac{\partial^2 \tilde{H}_1}{\partial z^2} = 4\pi N \, \mathrm{Sp}\!\left(\hat{\rho}\dfrac{\partial \hat{H}_P}{\partial \tilde{H}_1}\right) \qquad (2b)$$

where $\rho_{\mu\mu}$ and $\rho_{\mu\nu}$ are the components of the quantum density matrix, $\rho_{\mu\nu} = \langle\zeta_\mu|\hat{\rho}|\zeta_\nu\rangle$; $\hat{\rho}$ is the density operator; $\hat{H}$ is the system Hamiltonian, $\hat{H} = \hat{H}_0 + \hat{H}_P + \hat{H}_S$; $\hat{H}_0$ is the static spin-hamiltonian, $\hat{H}_0|\zeta_\mu\rangle = E_\mu|\zeta_\mu\rangle$; $|\zeta_\mu\rangle$ is the wave eigenfunction, belonging to energy level $E_\mu$; $\hat{H}_P$ and $\hat{H}_S$ represent pump and signal parts of system Hamiltonian dynamical component; transverse relaxation parameters $\tau_{2,\mu\nu} \equiv \tau_{2,\nu\mu}$ are the spin-spin relaxation times; longitudinal relaxation parameters $W_{\mu\nu} \neq W_{\nu\mu}$ are the probabilities of spin-lattice relaxation for the spin transitions $\zeta_\mu \leftrightarrow \zeta_\nu$ and $\zeta_\nu \leftrightarrow \zeta_\mu$ accordingly; $\tilde{H}_1$ is the amplitude of magnetic component of pump field inside of MFPR; $V_e$ is the light velocity in the crystal; $N$ is impurity paramagnetic centers concentration; $z$ is the coordinate along MFPR axis. In order to evaluate the inversion ratio the system (2) may be restricted by the wave equation for nonsaturating (probe) signal field, but this field do not change the population differences in the system by definition.

For our system the double inequality $\langle H_0 \rangle \gg \langle H_P \rangle \gg \langle H_S \rangle$ is valid and we will not take in account a) any spin-level energy renormalization on the pump transition; b) nonlinear phenomena on the probe field frequency. So it is the straightforward possibility to reduce complicated system of equations (2) to much more simple flow system using the same arguments as in [1-3, 8]. For one-frequency (e.g. three-level) model of pumping of quantum amplifier [5], using slowly varying pump field amplitude and phase approximation and McCall boundary conditions [1] for Bonifacio-Lugiato mean-field model of nonlinearity [11] we receive the following flow-type equations of motion for MFPR with 5 dynamical parameters (normalized components of vectorial order parameter):

$$\left. \begin{array}{rcl} d\tilde{B}_1/dt' &=& -\left[(\xi_0/2)\tilde{M}_1 + \tilde{B}_1/q_2 - \varphi_0\tilde{B}_2 - (P/q_2)^{1/2}\right]; \\ d\tilde{B}_2/dt' &=& -\left[(\xi_0/2)\tilde{M}_2 + \varphi_0\tilde{B}_1 + \tilde{B}_2/q_2\right]; \\ d\tilde{M}_1/dt' &=& -\left[\tilde{M}_1 - (h+\varphi_0)\tilde{M}_2 - \tilde{B}_1\tilde{D}_p\right]; \\ d\tilde{M}_2/dt' &=& -\left[(h+\varphi_0)\tilde{M}_1 + \tilde{M}_2 - \tilde{B}_2\tilde{D}_p\right]; \\ d\tilde{D}_p/dt' &=& -r_{21}\left[(\tilde{M}_1\tilde{B}_1 + \tilde{M}_2\tilde{B}_2) + \tilde{D}_p - 1\right], \end{array} \right\} \qquad (3)$$

where $t' = t/\tau_{2p}$; $\widetilde{B}_1 = \frac{1}{2\sqrt{r_{21}}} \text{Re}[\widetilde{\omega}_1^{(c)}(t')]$; $\widetilde{B}_2 = \frac{1}{2\sqrt{r_{21}}} \text{Im}[\widetilde{\omega}_1^{(c)}(t')]$; $\widetilde{\omega}_1^{(c)}$ — complex Rabi frequency [5] in MFPR; $\widetilde{M}_1 = \text{Re}\,\widetilde{M}$; $\widetilde{M}_2 = \text{Im}\,\widetilde{M}$; $\widetilde{M} = i\sqrt{2/q_2}\{[\widetilde{\rho}_{13}^{(RCS)}(t)]/\Delta\rho^{(0)}(\eta)\}$ — dynamical part of magnetic dipole moment in rotating coordinate system (*RCS*) ; $\varphi_0 = (\omega_c - \Omega)\tau_{2p}$; $h = [\omega_p(\vec{H}) - \omega_c]\tau_{2p}$; $\Omega$ — pump source frequency; $\vec{H}$ — static magnetic field; $\omega_p = \omega_p(\vec{H})$ — resonant frequency of spin pump transition $E_1 \leftrightarrow E_3$; $\omega_c$ — frequency of MFPR operation mode, $\Delta\rho^{(0)}(\eta)$ — thermodynamically equilibrium value of pump transition population difference; $\eta = \hbar\omega_p/(2S+1)k_B T$; $S$ — effective spin (equals to 1 for our three-level system); $k_B$ — Boltzmann constant; $T$ — temperature of thermostate; $\xi_0 = \pi f_C \hbar \omega_p \gamma_p^2 N \tau_{1p} \tau_{2p} \eta$; $\gamma_p$ is the effective hyromagnetic ratio for pump spin transition; $f_C$ is MFPR filling factor.

Thus we have at least 7 dimensionless control parameters for our reduced system of 5 equations of motions, namely $\xi_0$, $r_{21}$, $q_2$, $P$, $\varphi_0$, $h$, $\eta$; and very many of possible dynamical scenarios will be going on in common case of our $\mathbf{R}^5 \otimes \mathbf{R}^7$ space. In opposition to precise knowledge of dimension of phase space $\mathbf{R}^5$ for dynamical parameters, the dimension of control parameter space is not so definite, because of impossibility to split control parameter set into collection of fully independent values. One can easily see that, e.g., $\xi_0$, $r_{21}$ and $\eta$ are the functions of temperature, both $\eta$ and $h$ are the functions of $\vec{H}$, and so on. Moreover, the normalization of order parameter components may be dramatically changed by redefinition of control parameter set to avoid dependence of normalizing values on *scanned* control parameters.

### 4. Nonequilibrium stationary states and their asymptotic stability for the scalar (one-dimensional) model of pumping

The flow equations (3) for 5-dimensional vectorial order parameter, conrolled by 7 external parameters need further reduction. The simplest case from the dynamical point of view is the scalar model: dimension of phase space is equal to 1. It is obviously that in the case $\tau_c \ll \tau_{1p}$ the leading dynamical parameter is $\widetilde{D}_p$, because $r_{21}$ remains much smaller than 1 by definition. Thus an adiabatic approach [12] is valid for reductuion of order parameter dimension. In the framework of this approach, using the methods of catastrophe theory [13, 14], we will construct the critical surface $P = P(q_2, \xi_0, h, \varphi_0)$ in control parameter space. Control parameter $r_{21}$ is eliminated in adiabatic approximation, and thermostat themperature will be fixed here for simplicity. Thus from (1), (3) we obtain for the single component of the order parameter $\mathbf{X} = \mathbf{X}^{[1]}$:

$$X_1^{[1]} = \widetilde{D}_p; \quad \Psi_1^{(L)} = (1 - \widetilde{D}_p)/\tau_{1p}; \quad \Psi_1^{(NL)} = \frac{P_X \widetilde{D}_p}{(\widetilde{D}_p^2 + e_1 \widetilde{D}_p + e_{02})\xi^2} \tag{4}$$

where $e_1 = 4(1 - \delta\varphi)/\xi$; $e_{02} = 4(1 + \varphi^2)(1 + \delta^2)/\xi^2$; $\varphi = q_2\varphi_0$; $\delta = h + \varphi_0$; and $P_X$ is independent on $\tau_{1p}$ pump power parameter, $P_X = 4Pq_2/\tau_{1p} \equiv 2(\tau_c\tau_{2p}/\tau_R)\gamma_p^2 H_{1p}^2$; $\tau_R$ is the delay time for electromagnetic signal in MFPR; $H_{1p}$ is the amplitude of the input pump field magnetic component.

Let us find dependence of stationary states of (4) on $q_2$. The stationary normalized population difference of pumped spin transition $D_p(q_2) \equiv \widetilde{D}_p^{(st)}(q_2)$ may be expressed in the form of two single-valued inverse functions of the following form

$$q_{2\pm}(D_p) = \qquad (5)$$

$$= \frac{2(1+\delta^2)}{D_p} \left\{ \frac{P}{1-D_p} - \xi_0 \pm \left[ \frac{P}{1-D_p}\left(\frac{P}{1-D_p} - 2\xi_0\right) - \left(\xi_0\delta - \frac{2(1+\delta^2)\varphi_0}{D_p}\right)^2 \right]^{1/2} \right\}^{-1},$$

It is easy to show, that expression for the single Lyapunov exponent $\lambda$ of the flow equation (4) has the form

$$\lambda = \frac{(D_p-1)D_p e_{02}\xi^2}{4\tau_{1p}q_2 P}\left[\frac{1}{D_p^3} + \frac{e_1-1}{e_{02}D_p} + \frac{2}{e_{02}}\right] = \frac{1-D_p}{\tau_{1p}P}\left(\frac{\partial D_p}{\partial P}\right)^{-1}. \qquad (6)$$

The only type of critical phenomena in such 1-dimensional dynamical system is the saddle-node bifurcations. The supercritical saddle-node bifurcation takes place when the system come in his multivalued state and the subcritical one takes place when the system leaves off from pointed state. Let us find the explicit formulae for the bifurcation surface as the functions $P$ on $\xi_0$, $q_2$, $\varphi_0$, $h$. From $\lambda = 0$, using condition $0 < D_p < 1$, we obtain

$$P_{1,2} = -q_2^{-1}\sum_{m=0}^{3}k_m^{(p)}\left[D_p^{(1,2)}(\xi_0, h, \varphi_0, q_2)\right]^{m-1}(2/E_2)^{m-1}, \qquad (7)$$

where

$$k_3^{(p)} = -(q_2\xi_0 E_2/4)^2 \; ; \quad k_2^{(p)} = \frac{q_2\xi_0 E_2}{2}\left(\frac{q_2\xi_0}{4} - E_1\right);$$

$$k_1^{(p)} = q_2\xi_0 E_1 - E_2 E_0 \; ; \quad k_0^{(p)} = 2E_0 ;$$

$$D_p^{(1,2)} = \pm(1/2)E_3 \sec(\Theta_1 \mp \pi/6) ;$$

$$\Theta_1 = \frac{1}{3}\left\{\left[\arccos\left(\frac{3\xi_0 E_3}{4q_2^{-1}E_1 - \xi_0}\right)\right] + \frac{\pi}{2}\right\} ; \qquad (8)$$

$$E_3 = \frac{1}{2}\left\{3E_2 E_0\left[\left(\frac{q_2\xi_0}{4} - E_1\right)\frac{q_2\xi_0}{4}\right]^{-1}\right\} ;$$

$$E_2 = 1 + \delta^2 \; ; \quad E_1 = 1 - \delta\varphi \; ; \quad E_0 = 1 + \varphi^2 ;$$

Expressions (7), (8) represent the rigorous description of critical surface topology for the scalar pump model in the framework of the Bonifacio-Lugiato nonlinearity mechanism. The remarkable pecularity of such model is full coincidence of this critical surface (i.e. system bifurcation surface) with the $D_p$-double-degeneracy surface. Consequently, from the *topological* point of view formulae (7), (8) give full information about structure of stationary state spectra: inside of critical surface there are two nodes and one saddle state (separatrice) in the phase space, and it is one node outside of this surface. In other words, surface (7), (8) is the well-defined boudary between bistable and monostable states of our dissipative system. But this approach, which is standard one for many optical tasks [1, 11, 12] and which is widely used in applications of catastrophe theory [13, 14], does not take into consideration any *metrical* properties both of phase space and control parameters space stratification. On the other hand, Lyapunov stability analysis does not give any information about stability of stationary states relatively to finite perturbations of control parameters. And last but not least is the common question for the dissipative systems — the structural stability of possible dynamical states. In the next section we study these issues in details.

### 5. Critical surface cross-sections and inversion ratio spectra analysis for the scalar model of pumping

The inversion ratio spectrum $K_i(q_2, P, ...)$ for the probe signal on $E_1 \leftrightarrow E_2$ spin transition of our three-level spin system may be derived in the same manner as for usual single-valued $K$ in the standard theory of quantum amplifiers [5, 6]. Let us add wave equation for optical-wavelength signal to the equations (2)

$$\frac{\partial^2 \widetilde{U}}{\partial t^2} - V_u^2 \frac{\partial^2 \widetilde{U}}{\partial x^2} = \frac{N}{\rho'} \operatorname{Sp}\left(\widehat{\rho} \frac{\partial \widehat{H}_S}{\partial \varepsilon_{xx}}\right) \qquad (9)$$

where $\widetilde{U}$ is longitudinal deformation of crystalline lattice in the direction $Ox$; $\rho'$ is the crystal density; $\varepsilon_{xx}$ is the component of elastic deformation tensor. After straightforward calculations the inversion ratio for the whole system of equations (2),(9) may be expressed in the form

$$K_j = \frac{(K_m + 1)\left[1 - D_p^{(j)}(q_2, P, \xi_0, h, \varphi_0)\right] - 1}{1 + b^2 h^2}; \quad j = 1, 2, 3; \qquad (10)$$

where $K_m$ is the limit value of inversion ratio [5], $b = (\tau_{2s}/\tau_{2p})(\partial \omega_s / \partial \omega_p)$ ; $\tau_{2s}$ is longitudinal spin relaxation time at signal transition frequency $\omega_s$; and $D_p$ is derived from (5).

Before analyzing inversion ratio spectra (10) let us consider the cross-sections of critical surface (7), (8) by the hyperplanes $\{\xi_0, h, \varphi_0\}$. The qualitative distinction between cross-sections having $\varphi_0 \neq 0$ and $\varphi_0 = 0$ is clearly seen from bifurcation diagrams in the $\{q_2, P\}$ plane (see formulae (7), (8) for fixed remnant of control parameters). In the first case, as $P$ increases the domain of multi-valued inversion ratio $K$ always remains limited (although not necessarily continuous) versus $q_2$. Conversely, with $\varphi_0 = 0$ as $P$ is increasing the usual hysteretical (on $q_2$) part of the $K(q_2)$ dependence will rapidly expands towards the greater and greater values of $q_2$ and finally is replaced by semibounded domain where the upper asymptotically stable branch $K_3^{(\varphi_0)}(q_2)$ is seen to coexist with an isolated asymptotically stable branch $K_1^{(\varphi_0)}(q_2)$. And the intermediaty branch $K_2^{(\varphi_0)}(q_2)$ due to inequality $\partial Z_p^{(j)}/\partial P < 0$ $\left(Z_p^{(j)} \equiv \frac{1 - D_p^{(j)}}{D_p^{(j)}} e_2\right)$ is always unstable and represents, as it was pointed earlier, the separatrix of attractors $^{(0)}\mathbf{A}_{\text{Low}}$ and $^{(0)}\mathbf{A}_{\text{High}}$ (stable nodes for our scalar model) corresponding to the branches $K_1^{(\varphi_0)}(q_2)$ and $K_3^{(\varphi_0)}(q_2)$ respectively.

It can be shown that the branch $K_3^{(\varphi_0)}(q_2)$ is the continuation of the single-valued function $K^{(\varphi_0)}(q_2)$ after the critical surface is crossed by the imaging point; and we have $\partial K^{(\varphi_0=0)}/\partial q_2 > 0$ throughout the entire range of $q_2$. This is in consistency with the naïve idea about mechanism of nonequilibrium state formation: the greater is the pump resonator quality (i.e. the greater is $q_2$), the higher is the inversion ratio under the fixed pump power. As far as the lower (normally noninverted) branch $K_1^{(\varphi_0)}(q_2)$ is concerning, it could seemingly be ignored (it is an isola [15]).

As a matter of fact, the situation appears to be more complicated. For instance, using the formulae (5), (10), we find that as $q_2$ increases the branch $K_3^{(\varphi_0)}(q_2)$ becomes as close as desired to the separatrice $K_2^{(\varphi_0)}(q_2)$. Hence, the small but finite perturbations would throw the imaging point out of the $^{(0)}\mathbf{A}_{\text{High}}$ attractor's bath, thereby switching the system from the strongly nonequilibrium state to the almost equilibrium one (where the inversion ratio has negative value). In other words, the inversion state of paramagnet under conditions of great values of $q_2$, which forms as the result of the synphase reemission of pump field in the MFPR, is consistent with a very shallow local minimum of the dissipative system potential [9–11]. Conversely, the noninversion state for the same values of $q_2$, corresponding to the antiphase reemission of pump field, is stable even against significant perturbations of the system stationary state: the electromagnetic pump energy is used not to provide acoustic (phonon) microwave induced emission at signal spin transition but to keep the absorbing state of paramagnet. Consequently, the simple scalar optical model of bistability [12] is not applicable for our pump system at least for $q_2 >> 1$ in spite of the inequality $\tau_c << \tau_{1p}$ is true for the nonlinear MFPR under cosideration.

Thus, if one aspires for an MFPR quality factor increasing to reduce the needed for inversion pump power he brings an adverse result — the inversion state collapse caused by the dangerous approaching of $^{(0)}\mathbf{A}_{\text{High}}$ to the separatrice. Moreover, the case of the resonator "precise tuning" $\varphi_0 = 0$ is the exception (singularity), because slightest detunings of the pump source frequncy from the MFPR eigen frequency transforms the above-mentioned semi-bounded domain of asymptotical critical surface cross-section to the bounded (may be discontinuous) one along the $q_2$. It is the manifestation of the structural instability of the inversion states formed under conditions of the pump field synphase reemission in high-quality MFPR.

The resonance form of the dependencies $K(q_2)$ at $\varphi_0 \neq 0$ is clearly seen from formulae (5), (10): the nonequilibrium steady state of the paramagnet is created, passes through the maximum value and then smoothly dissapiers or sharply destroys in sertain interval (intervals) of values of $q_2$. The above-mentioned nonlinear resonance is characterized by singlevalued or multivalued dependence of $K$ upon $q_2$, and for a certain interval of values $q_2$ the inversion asymptotically stable stationary state is lying very close to the separatrice (intermediaty branch in (10), i.e. $K_2$) and, in addition to this, it has the isola, where the supercritical and subcritical saddle-node bifurcations are corresponding to the same pair of inversion ratio branches [2, 15]. The last circumstance means that it is no possibility for any soft excitation of inversion state in the isola domain on the $q_2$ axis. Therefore, the real width of such resonance obseved via continuous one-dimensional scanning of the control parameter is considerably less, then its value specified from the distance between limiting points of the whole domain of $q_2$ contained all the intervals having $K \geq 0$. The knowledge of critical surface cross-sections (i.e. understanding only the topological structure of phase space stratification) does not give such very importatnt information. The correct choice of pumping system parameters must include metrical analysis of properties for each attractor to achieve stable nonequilibrium state with a maximum possible inversion ratio.

### 6. Vectorial (3D) model of pumping: instabilities and the critical surfaces

If the MFPR quality factor is high, i.e. $\tau_c \gg \tau_{2p}$, especially if $\tau_c$ is of order of $\tau_{1p}$, the critical phenomena in pump system are very rich due to extension of phase space dimension [17], [18]. In this case the parameter of order for the dissipative system becomes vectorial and contains three components. Introducing $t'' = t/2\tau_c$, we can reduce (3) to the 3D set of differential equations, yelding the flow:

$$\frac{d\mathbf{X}^{[3]}}{dt''} = \Psi^{(L)}(\mathbf{X}^{[3]}, \mathbf{C}) + \Psi^{(NL)}(\mathbf{X}^{[3]}, \mathbf{C}), \qquad (11)$$

where $\mathbf{X}^{[3]} = \{\tilde{B}_1; \tilde{B}_2; \tilde{D}_p\}$ ; $\mathbf{C} = \{A; q_1; q_2; Y; \varphi; h; \eta\}$. The control parameters vector $\mathbf{C}$ is slightly redefined here ($A \equiv \xi/2$; $Y \equiv \sqrt{Pq_2}$). The components of $\mathbf{X}^{[3]}$, $\Psi^{(L)}$ and $\Psi^{(NL)}$ are:

$$\left. \begin{array}{lll} X_1^{[3]} = \tilde{B}_1; & \Psi_1^{(L)} = Y - \tilde{B}_1 + \varphi\tilde{B}_2; & \Psi_1^{(NL)} = -\dfrac{A}{1+h^2}(\tilde{B}_1 + h\tilde{B}_2)\tilde{D}_p; \\[2mm] X_2^{[3]} = \tilde{B}_2; & \Psi_2^{(L)} = -\varphi\tilde{B}_1 - \tilde{B}_2; & \Psi_2^{(NL)} = \dfrac{A}{1+h^2}(h\tilde{B}_1 - \tilde{B}_2)\tilde{D}_p; \\[2mm] X_3^{[3]} = \tilde{D}_p; & \Psi_3^{(L)} = (1-\tilde{D}_p)q_1; & \Psi_3^{(NL)} = \dfrac{q_1}{1+h^2}(\tilde{B}_1^2 + \tilde{B}_2^2)\tilde{D}_p; \end{array} \right\} \qquad (12)$$

Temperature of thermostate is fixed, so $\eta$ is an hidden control parameter (as it was in scalar model too). The magnetic field dependencies (i.e. $h$-dependencies) are the main object of our studies for 3D model of pumping. These dependencies were not investigated in works on optical bistability [17], [18].

Stationary states spectrum of flow (11), (12) in the common case lies on the Whitney cusp surface $W(x \mid u, v) \equiv x^3 + ux + v = 0$ [14], which for our case has the following form:

$$W_h(D_p) = W(D_p - g_h \mid u_h, v_h) = 0, \qquad (13)$$

where

$$g_h = \frac{1}{3}(1 - 2f_1); \quad u_h = -\frac{1}{3}[1 + 2f_1(1 + 2f_1)] + f_{02} + f_3;$$

$$v_h = (u_h + g_h^2)g_h - f_{02}; \quad f_k = [1 + (-1)^k h^k \varphi^{2-k}]A^{-k}; \qquad (14)$$

$$f_{02} = f_0 f_2; \quad f_3 = Y^2/A^2; \quad k = 0, 1, 2.$$

In the same manner as for scalar model, we may construct the critical surface which separate singlevalued and multivalued stationary states of pump system. This surface is the geometrical place of the double-degenerate points $^{II}D_p$ in the control parameters space, and the analytical expression for this *Type I critical surface* is the same as (7), (8) after the obvious replacement $\delta \to h$.

At this point the similarity of $\mathbf{X}^{[1]}$ and $\mathbf{X}^{[3]}$ dynamical states is ended. Let us find the Lyapunov stability of the $D_p^{(i)}$ branches $(i = 1, 2, 3)$ by linearizing the 3D flow equations (11) and constructing the Hurwitz determinants $S_1, S_2, S_3$ [16] for the characteristic equation:

$$S_1 = \zeta + \Pi_1 \; ; \quad S_2 = \Pi_2 \zeta^2 + (\Pi_{12} + \Pi_3)\zeta + \Pi_{01} \; ; \quad S_3 = \frac{\zeta S_2}{(Z_p)'} , \tag{15}$$

where

$$\zeta = q_1 / D_p(h) \; ; \quad \Pi_{ab} = \Pi_a \Pi_b \; ; \quad \Pi_0 = Y^2 / Z_p \; ; \quad \Pi_1 = 2 + \Pi_4 \; ;$$

$$\Pi_2 = 2 + \Pi_4 D_p(h) \; ; \quad \Pi_3 = \Pi_0 - \frac{1}{(Z_p)'} \; ; \tag{16}$$

$$\Pi_4 = 2 D_p(h) / A f_2 \; ; \quad (Z_p)' = \frac{\partial Z_p}{\partial (Y^2)} ,$$

and the $D_p^{(j)}(h)$ may be found from the pair of inverse dependencies:

$$h_\pm(D_p) = \left(\frac{AD_p}{f_0} - \frac{1}{q_2}\right)\varphi \pm \left\{\frac{Y^2 D_p}{(1 - D_p) f_0} - \left[\frac{AD_p}{f_0} + 1\right]^2\right\}^{1/2} . \tag{17}$$

The asymptotic stability criteria are: $S_1 > 0$; $S_2 > 0$; $S_3 > 0$. It is always $\text{sgn}(Z_p)' = -1$ at the second branch, i.e. it is asymptotically unstable throughout the whole domain of multivalued pump states. The boundaries of asymptotic stability for the first and the third branches in the pointed domain or for their continuations in the domains of singlevalued states are defined by the conditions $S_2^{(1,3)} = 0$, because of equality $\text{sgn}(Z_p)' = +1$ at these branches ( and it is always $S_1 > 0$ ). The surfaces

$$S_2^{(1,3)}(\mathbf{C}) = 0 \tag{18}$$

are *Type 2 critical surfaces* for our dissipative system under 3D vectorial model of pumping.

For $q_1 \gg 1$ or $q_1 \ll 1$ branches $D_p^{(1)}$ and $D_p^{(3)}$ are asymptotically stable and both critical surfaces coincide similarly to the scalar model of pumping. But Lyapunov instability takes place at least for one of these branches in the case $q_1^{(B-)} < q_1 < q_1^{B(+)}$, where $S_2(\mathbf{C}) < 0$, i.e. when

$$(Z_p)' < \left(\Pi_{12}^{1/2} + Z^{-1/2} Y\right)^{-2} . \tag{19}$$

Here new bifurcation parameters $q_1^{(B\pm)}$ have the form for every branch:

$$q_1^{(B\pm)} = \left(-r_a \pm \sqrt{r_a^2 - \Pi_{01} \Pi_2^{-1}}\right) D_p \; ; \quad r_a = 1 + \frac{1}{2}\left(\Pi_4 + \Pi_3 \Pi_2^{-1}\right) . \tag{20}$$

Finally, the new critical surfaces have been found for a 3D vectorial model of pumping of three-level paramagnetic system with high-quality resonator. Not only structural instability and instabilities caused by finite perturbations, but the Lyapunov instability is possible for high-quality resonator at the upper branch of a bistable pump system. The above results allows us to form a fully analytical picture of the physical mechanisms that may lead to instabilities of bistable pump resonators and phaser pump resonators in particular. These mechanisms are also of great interest for interpretaion of some experimental data on 1-μm ruby phaser [19–22]. Note that high-quality whispering-gallery resonator in the pump system of a solid-state maser generator (a cryogenic sapphire frequency standard) [23–25] is still below the threshold of the instability described by us in Sec. 6 because of *very low concentation* of the active resonant paramagnetic centers $Fe^{3+}$, so the magnetic quality $Q_m$ (which is inversely proportional to the concentration of paramagnetic centers) is not low. As the result. the ratio $Q_c/Q_m$ in [23–25] is not as high as it follows from (19), (20). On the other hand, *bistability itself was observed* in [24–25] (at the signal transition frequency). So, at higher concentrations of $Fe^{3+}$, conditions (19), (20) may be fulfilled for the system describes in [23–25] too.